\documentclass[10pt,letterpaper]{article}

\usepackage[latin1]{inputenc}
\usepackage{amsmath}
\usepackage{amsfonts}
\usepackage{amssymb}
\usepackage{graphicx}
\usepackage{fullpage}
\usepackage{float}
\usepackage{multicol,caption}
\usepackage[parfill]{parskip} 
\usepackage{color}
\usepackage[colorlinks=true, pdfstartview=FitV, linkcolor=blue,citecolor=blue, urlcolor=blue]{hyperref}
\usepackage{osajnl2}
\usepackage{epstopdf}
\usepackage{booktabs}
\usepackage{ctable}
\usepackage{array}
\usepackage{subfig}
\usepackage{cite}

\newcommand{\mytab}{
\centering
\begin{tabular}{lcccccc}
\toprule
					& $R_{0}$		& $\lambda_{cutoff}$	& $NA$	& $V_{core,R_0}$	& $R_c^{LP_{11}}$	& $V_{clad,R_c}$\\
					& ($\mu$m)	& (nm)				&		&				& ($\mu$m)		&	\\\otoprule
Thorlabs SM980G80		& 40			& 920				& 0.18	& 2.9				& 33				& 288\\
Fibercore SM1500		& 25			& 1396				& 0.3		& 4.3				& 14			& 119\\\bottomrule
\end{tabular}
}

\begin{document}

\title{A low-loss photonic silica nanofiber for higher-order modes}

\author{S. Ravets,$^{1,2}$ J. E. Hoffman,$^{1}$ L. A. Orozco,$^{1,*}$ S. L. Rolston,$^{1}$ G. Beadie,$^3$ and F. K. Fatemi$^3$}

\address{$^{1}$Joint Quantum Institute, Department of Physics, University of Maryland and National Institute of Standards and Technology, College Park, Maryland 20742, USA.\\
$^{2}$Laboratoire Charles Fabry, Institut d'Optique, CNRS, Univ Paris Sud,
2 Avenue Augustin Fresnel, 91127 Palaiseau cedex, France.\\
$^{3}$Optical Sciences Division, Naval Research Laboratory, Washington DC, 20375, USA.\\
$^{*}$lorozco@umd.edu}


\begin{abstract*}
Optical nanofibers confine light to subwavelength scales, and are of interest for the design, integration, and interconnection of nanophotonic devices. Here we demonstrate high transmission ($>$ 97\%) of the first family of excited modes through a 350~nm radius fiber, by appropriate choice of the fiber and precise control of the taper geometry. We can design the nanofibers so that these modes propagate with most of their energy outside the waist region. We also present an optical setup for selectively launching these modes with less than 1\% fundamental mode contamination. Our experimental results are in good agreement with simulations of the propagation. Multimode optical nanofibers expand the photonic toolbox, and may aid in the realization of a fully integrated nanoscale device for communication science, laser science or other sensing applications.
\end{abstract*}

\section{INTRODUCTION}
\label{sec:intro}

Optical nanofibers with a waist diameter smaller than the wavelength of the guided light are currently used for non-linear optics, atomic physics, sensing, and fiber coupling \cite{Kien2004,Leon-Saval2004,Vetsch2010}. The intense evanescent field outside of an optical nanofiber is particularly interesting for atom trapping and strong atom-photon coupling. Tapering standard optical fiber down to submicron diameters has been a successful fabrication technique for a variety of applications \cite{Tong2003,Birks1992,Jiang2006,Nayak2011,Wuttke2012}, in which transmissions of the fundamental mode of the waveguide reaching more than 99\% have been achieved \cite{Tong2003, Brambilla2004}. Until now, the efficient guidance of higher-order modes has not been observed, due to the ease with which they couple to other modes, leading to large losses. This restricts most work with nanofibers to the single mode regime, where the diameter is small enough to only support the fundamental mode $HE_{11}$ \cite{Vetsch2010,Spillane2003}.

This paper reports measurements with nanofibers using the first excited $TE_{01}$, $TM_{01}$, and $HE_{21}$ modes, which have azimuthal, radial, and hybrid polarization states, respectively. Prior work on higher-order mode propagation has shown $\approx$30\% transmission of this $LP_{11}$ family of modes~\cite{Frawley2012}, with a mode purity at the fiber output of $\approx$70\%, which gives a total transmission of only 20\%. Here, by carefully controlling the taper geometry, and by choosing a commercially available 50 micron reduced-cladding-diameter fiber with increased numerical aperture, we demonstrate transmissions greater than $97\%$ for light of 780~nm wavelength in the first excited $LP_{11}$ family of modes through fibers with a 350 nm waist radius. Furthermore, we present a setup that enables us to efficiently launch these three modes with exceedingly high purity at the output, where less than $1\%$ of the light is coupled to the fundamental mode \cite{Fatemi2011}. We follow the work of \cite{Warken2007b} to fabricate the nanofibers and use a series of diagnostics during the pull to monitor the quality of the fiber. We record the total transmission of the light and image the mode exiting the fiber for the duration of the pull. Analysis of the transmission as a function of time for different types of fibers allows us to estimate which modes are excited during the pull as well as their relative energies of excitation through the use of spectrograms \cite{Ravets2013}.

These results open the way to efficiently use higher-order modes in optical nanofibers. Unlike the $HE_{11}$ mode, higher-order modes experience a cutoff at a finite radius. This allows improved control of the evanescent field extent at large radii, enabling stronger fibers and improved handling characteristics. Our work enables the usage of higher-order modes for atomic physics applications. In particular, the spatial interference between several of those modes can create unique evanescent field distributions on the waist, providing an easy and self-consistent way to break the symmetry along the propagation axis, suppressing the need to create a standing wave. This is particularly relevant to atomic physics applications for the realization of a one color, blue-detuned and state insensitive trapping potential for atoms \cite{Sague2008}.

\section{MODE PROPAGATION IN AN OPTICAL FIBER}
\label{sec:modes}

References \cite{Snyder1983,Yariv1990} describe the modes in a cylindrical waveguide using Maxwell equations. The modal fields vary as $\exp{\left[i(\beta z-\omega t)\right]}$ where $\beta$ is the propagation constant of the mode. A mode propagates in the fiber with an effective index 

\begin{equation}
n_{eff} = \beta/k,
\end{equation}
where $k=2\pi/\lambda$ is the free-space propagation constant and $\lambda$ is the free-space wavelength of the light. The propagation of light inside a two-layered step-index fiber, consisting of a core of radius $a$ and refractive index $n_{core}$ surrounded by a cladding of radius $R$ and refractive index $n_{clad}$, depends on $V$, 

\begin{equation}
V=\frac{2\pi a}{\lambda}\sqrt{n_{core}^2-n_{clad}^2}.
\label{V}
\end{equation}

$V$ plays an important role in our tapers, since the radius, $a$ in Eq.~(\ref{V}), varies enough that the interfaces seen by the modes change as they propagate through the taper.
At the beginning of the taper, the light is confined in the core, and guided by the core-to-cladding interface with $V_{core} $ as in Eq.~(\ref{V}), and $n_{clad} < n_{eff} < n_{core}$.
At the end of the taper, the core is negligible ($a_{core} \approx $10 nm $ \ll \lambda$) and the light is guided by the cladding-to-air interface with $n_{air} < n_{eff} < n_{clad}$.
Between these two regimes, the light escapes the core to the cladding, and the relevant radius in Eq.~(\ref{V}) becomes $R$, which is much larger than $a$. Due to this radius increase, and the large index difference between $n_{clad}$ and $n_{air}$, $V>>1$. The fiber becomes highly multimode, as the number of bound modes is proportional to $V^2/2$ \cite{Snyder1983}. Maintaining adiabaticity through this transition is critical.

\subsection{First family of excited modes}
\begin{figure}[H]
\centering
\includegraphics[width=10cm]{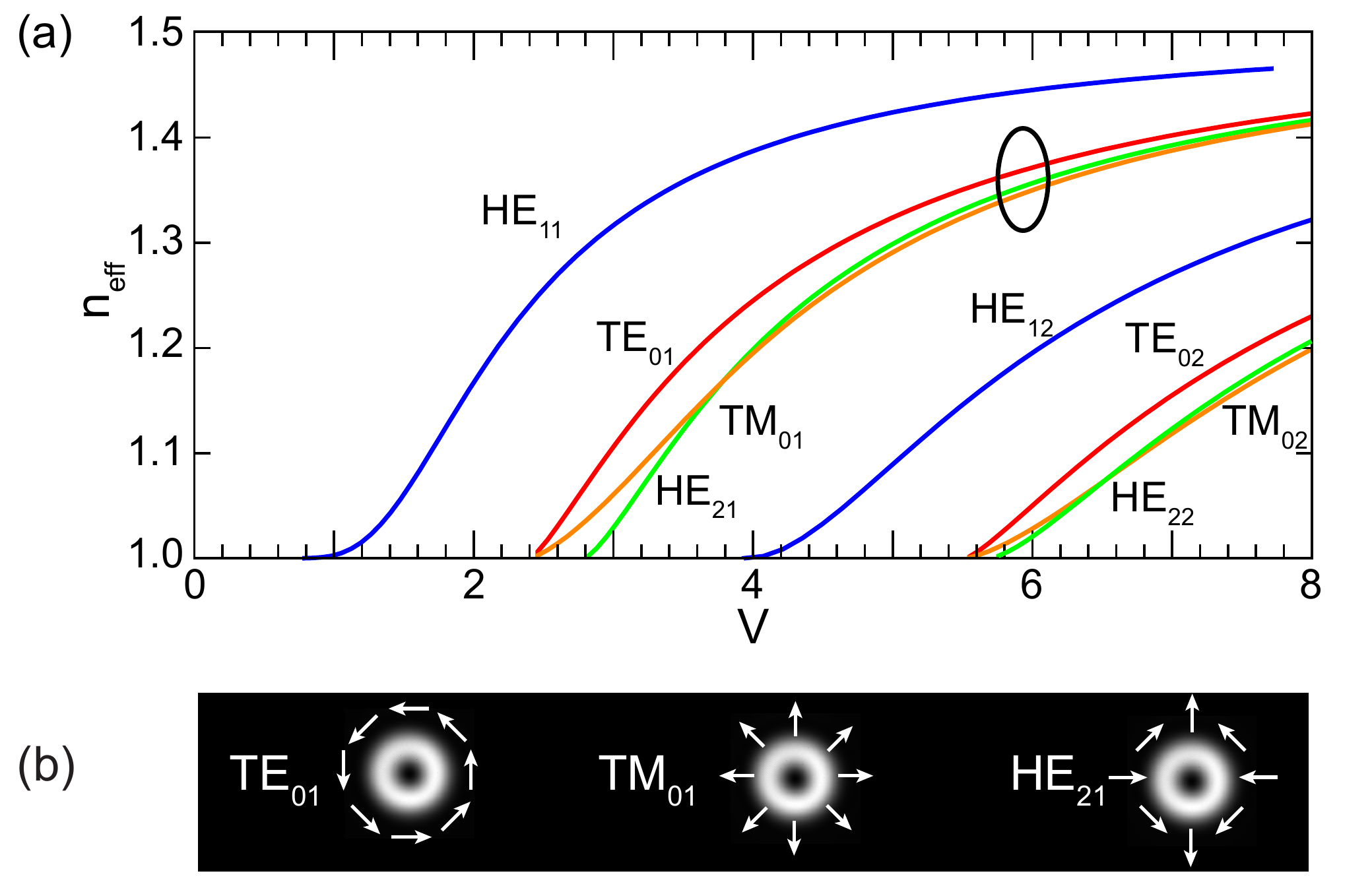}
\caption{ (a) $n_{eff}$ indices of several low-order modes in a nanofiber with $n_{clad}=1.5$, surrounded by vacuum ($n_{air}=1$). Below $V \approx 3.8$, two families of modes exist. In this work, we are emphasizing the $LP_{11}$ family (circled). By symmetry, modes in this family can interfere with the $TM_{02},\ TE_{02},\ HE_{22}$ family, which may be excited through non-adiabatic processes. (b) Intensity and polarization profiles of the $LP_{11}$ family of modes considered in this work.}
\label{modes}
\end{figure}

Figure~\ref{modes}(a) shows the effective index of refraction against $V$ for several low-order modes in a nanofiber. When $V < 2.405$, the fiber supports only the $HE_{11}$ mode. For a typical nanofiber, $n_{core} \approx 1.5$ and $n_{clad}=1$. The $TE_{01}$ and $TM_{01}$ modes are allowed for $V > 2.405$, and the $HE_{21}$ mode is allowed for $V > 2.8$. As long as $V$ remains lower than 3.8, only the four modes mentioned above are allowed. In the weakly guided regime, the modes of interest are known as the $LP_{11}$ family. We initially launch into the $actual$ $LP_{11}$ family, because at the entrance, the fiber is weakly guided. The taper takes us into the strong guiding regime where the $LP_{11}$ splits into $TE_{01}$, $TM_{01}$, and $HE_{21}$. We will refer to the set of modes ($TE_{01}$, $TM_{01}$, and $HE_{21}$) during the entirety of the pull as the $LP_{11}$ family. This simplifies discussions when referring to the full set of modes, especially in reference to excitations to modes or families with the same symmetry, i.e. $LP_{12}$ for $TE_{02}$, $TM_{02}$, and $HE_{22}$. Prior work has emphasized propagation of the $HE_{11}$ mode, where $V \leq 2.405$~\cite{Nayak2011,Vetsch2010, Spillane2003}. We are interested in selectively exciting and guiding the LP$_{11}$ family through a nanofiber, in a regime where $2.405 \leq V \leq 3.8$. When expanded to free space, these modes have the intensity and polarization profiles shown in Fig.~\ref{modes}(b).

\subsection{Adiabaticity condition}
\label{sec:adiabaticity}

During the pull we continuously decrease $R$ and the light escapes the core to the cladding as it propagates through the taper. After light escapes the core, the presence of the core, the cladding, and the air influence the mode. The fiber is highly multimode, and modes of the same symmetry can couple to each other. The mode evolution in a taper is strongly related to the shape of the taper. If a taper is too steep, the mode evolution is non-adiabatic resulting in low transmission. As the tapering angle $\Omega$ is reduced, the mode propagation becomes more adiabatic. Following this reasoning, an adiabaticity criterion has been derived \cite{Snyder1983} relating the characteristic length of the taper $z_t$, to the characteristic beat length between two modes $z_b$ where $z_t ={R}/{\tan(\Omega)}$, and $z_b ={2 \pi}/{(\beta_1 - \beta_2)} = {\lambda}/{(n_{{eff}_1}-n_{{eff}_2})}$.

The mode evolution in a taper is adiabatic when the fiber is long enough to satisfy $z_t \gg z_b$ \cite{Snyder1983}. For a typical silica optical fiber single mode at $\lambda = 780$~nm, we calculate using this criterion a minimum $\Omega$ of a few milliradian for the limiting case of $z_t =z_b$. This implies that a taper requires sub-milliradian $\Omega$ to achieve adiabaticity in the region where light leaves the fiber core and becomes a cladding mode. Therefore, choosing an optical fiber with a small initial cladding radius (and thus a reduced $V_{clad}$) combined with a high numerical aperture is highly advantageous to maintaining adiabaticity~\cite{Frawley2012}. Additionally, such a fiber reduces the overall drawing time, length requirements of the pulling apparatus, and the overall length of the taper.

\section{EXPERIMENTAL SETUP}

\begin{figure}[H]
\centering
\includegraphics[width=10cm]{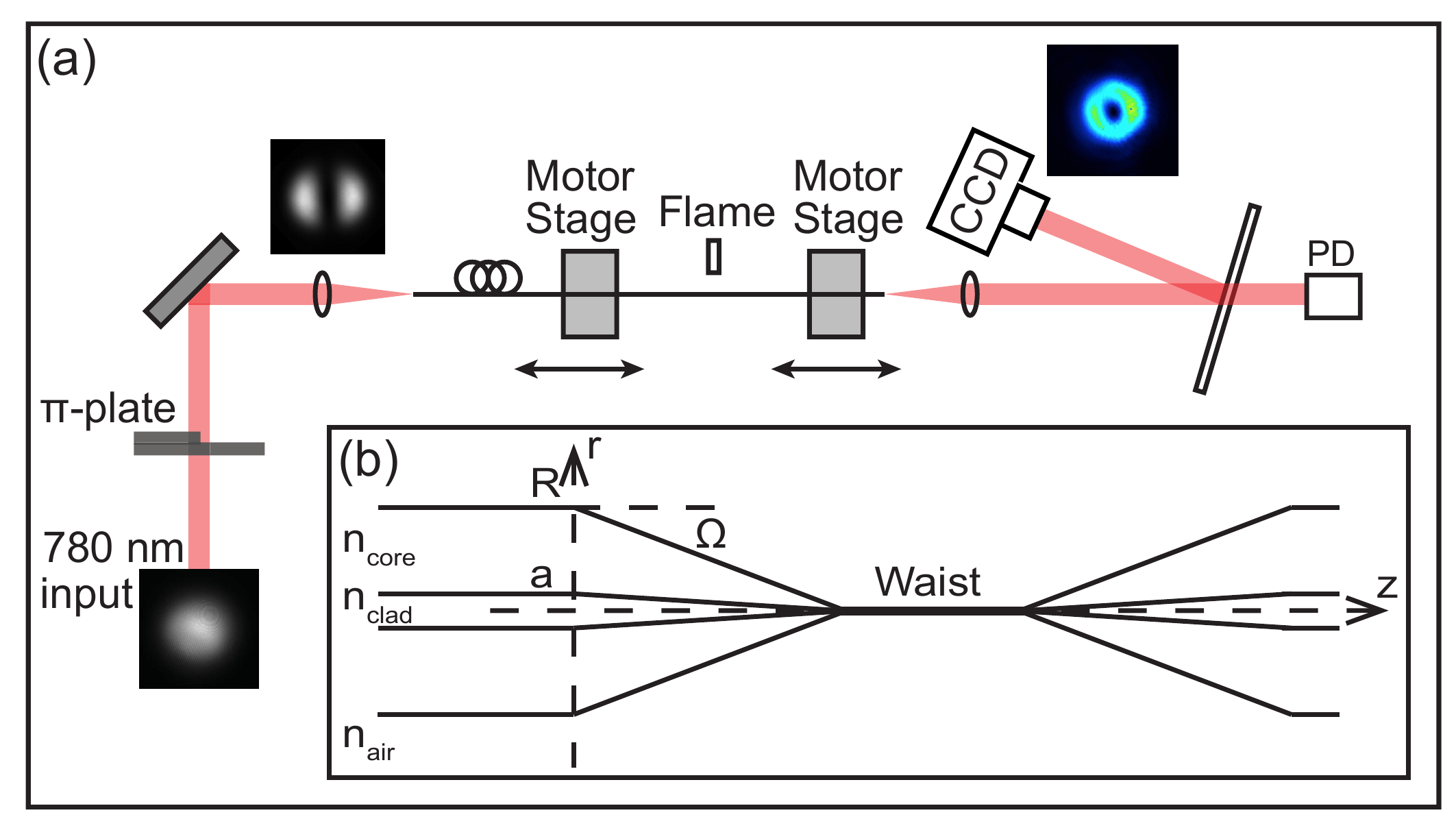}
\caption{ (a) Simplified diagram of the experimental setup. A Gaussian beam passes through a $\pi$-phase plate and is coupled into the fiber to be drawn. On the output of the fiber, a photodetector (PD) and camera (CCD) monitor the transmission. Typical beam images are shown. (b) Schematic of tapered nanofiber.}
\label{setup}
\end{figure}

Figure~\ref{setup} shows a diagram of our experimental setup. Here, we review the mode preparation, the pulling process, and the detection and analysis. We efficiently excite the $TM_{01}$, $TE_{01}$, and $HE_{21}$ modes using a fiber-based Cylindrical Vector Beam (CVB) generation method~\cite{Fatemi2011, Pechkis2012}. The fiber is drawn using a heat-and-pull method \cite{Warken2007b, Dimmick1999,Hoffman2013}. We measure the transmission while pulling, recording simultaneously the output of the photodiode to a digital storage oscilloscope and beam profiles on a Charge-Coupled Device (CCD) camera. 

A New Focus Vortex laser delivers a $TEM_{00}$ Gaussian beam at $\lambda$=780.24 nm. This wavelength corresponds to the $D2$ line of Rubidium, our atom of interest. We spatially filter this beam using a polarization-maintaining optical fiber, and collimate with an asphere to a 1/$e^2$ diameter of 630 microns. The Gaussian beam passes through a phase plate that imparts a $\pi$ phase shift on half of the beam, producing a two-lobed beam that approximates a $TEM_{01}$ free-space optical mode. The phase plate is on a translation stage, so that we can easily switch between the $LP_{11}$ and $LP_{01}$ families without adjusting alignment. Figure~\ref{setup} shows profiles of the beam before and after the phase plate. We couple the beam into the fiber using a matched asphere. The coupling coefficients to the fiber modes are determined by the beam symmetry. The inversion of the polarization over half the incident beam allows us to selectively excite the $LP_{11}$ family with high efficiency. Stress-induced birefringence at the input end of the fiber can be used to selectvely excite the $TM_{01}$, $TE_{01}$, and $HE_{21}$ modes individually \cite{Fatemi2011}, but we have not used that selectivity here. The fundamental $HE_{11}$ mode is only excited through aberrations in the beam. To achieve efficient coupling into the $LP_{11}$ family, we wrap the fiber with two or three windings around a 4-mm diameter mandrel that attenuates any modes higher than the $LP_{11}$ family. Figure~\ref{setup}(a) shows typical input (black and white) and output (color) modes.

We detail our nanofiber pulling process in Ref.~\cite{Ravets2013,Hoffman2013}. We start by clamping a fiber to two high-precision Newport XML 210 motor stages. We strip the buffer from the fiber and clean it thoroughly following the procedures in \cite{Hoffman2013}. We image the fiber with an optical microscope and remove any undesired particle remaining on the fiber surface. This is important for the repeatability of our measurements. An oxyhydrogen flame with a stoichiometric combination brings a 0.75 mm long portion of the fiber to a temperature that exceeds the softening point of fused silica. We pull on the fiber ends at a typical relative velocity of 0.1 mm/s. We calculate the stage trajectories to produce a fiber of a chosen geometry using an algorithm
that relies on conservation of volume \cite{Birks1992}. We divide the pull into approximately 100 steps. Each step adds a small section to the taper that reduces the radius of the waist. The sections are small enough to be considered linear locally. The compilation of each small taper creates the final taper with a desired geometry, which is generally composed of a few mrad steep linear section that reduces to a radius of 6 $\mathrm{\mu m}$, and then connects to an exponential section that gradually reaches a submicrometer radius. The central region forms a 7-mm uniform waist (see Fig.~\ref{setup}(b) for the geometry). To observe all possible mode cutoffs, the fibers are typically drawn to $R\approx$280 nm, at which point only the fundamental $HE_{11}$ mode propagates. We vary $\Omega$ from 0.4 mrad to 4 mrad, resulting in pull times lasting between 100 to 1000 seconds
The output side of the fiber is held straight with no mode filtering. We follow the mode evolution for the duration of the pull by monitoring the transmission of a few mW of laser power through the fiber. The transmitted beam is monitored both by a CCD and by a photodetector (PD). The beamsplitter shown in the figure is tilted to as small an angle as possible to eliminate polarization-dependent reflections. Another PD, not shown in Fig.~\ref{setup}(a), measures the input laser power during the pull to normalize the transmission signal. We record PD signals with a Tektronix DPO7054 oscilloscope with 16-bit resolution and a sample rate of 1-10 ksamples/s.

\section{RESULTS}

In this section, we present our results on the transmission of the $LP_{11}$ mode family in the context of adiabaticity. We first analyze the improvements gained by choosing a fiber with a reduced cladding radius and increased numerical aperture. In \cite{Frawley2012}, Frawley et al. looked at the improvement obtained by reducing the initial $R$ from 125~$\mu$m to 80~$\mu$m. Moving from 80~$\mu$m to 50~$\mu$m fibers, we improve the transmission from 10\% to 51\% for a 2-mrad taper. A spectrogram analysis depicts fewer and weaker excitations of higher-order modes. Second, we demonstrate an improvement in adiabaticity through the control of the taper geometry. Our ability to vary $\Omega$ allows for marked improvement in guidance efficiency through the nanofiber, increasing from 16\% at 4 mrad up to 97.8\% at 0.4 mrad.

\subsection{Varying the Fiber Type}

\begin{figure}[H] 
\captionsetup[subfigure]{labelformat=empty}
\centering
\includegraphics[width=10cm]{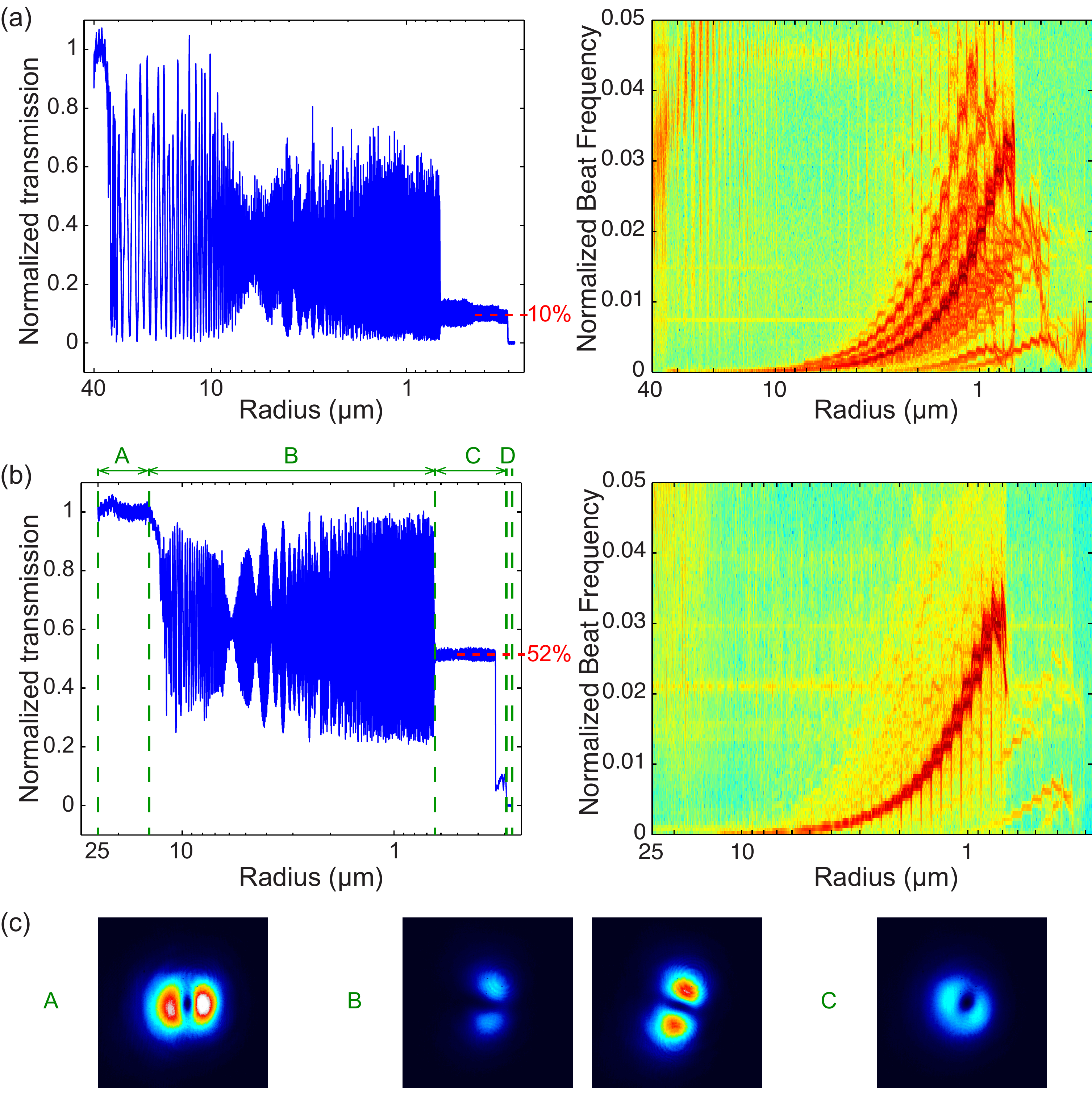}
\subfloat[ ]{\mytab}
\caption{Evolution while tapering of the transmission through fibers with a half angle of 2 mrad as a function of the radius of the waist. (a) Fiber with an initial diameter of 80~$\mu$m. (b) Fiber with an initial diameter of 50~$\mu$m. The spectrograms associated with those transmission curves give a clear picture of the power transfers during the pull. Note the logarithmic scales on the horizontal axis. (c) (1.0 MB) Movie of the evolution of the beam transmitted through the fiber measured on the CCD during a 2 mrad pull of the 50~$\mu$m fiber. We record one frame every second, and display them at a 7 frames per second speed. Sections A, B, C, and D are described in the text. The properties of the fibers used are summarized in the table.}
\label{fibertype}
\end{figure}

\subsubsection{Transmission measurements}

Figure~\ref{fibertype} shows typical transmissions obtained when exciting the fiber with a combination of $LP_{11}$ modes. We plot the normalized transmission during a single pull as a function of the fiber waist radius. We can identify four distinct regimes in Fig.~\ref{fibertype}. The modes are initially confined to the core (regime $A$ in Fig.~\ref{fibertype}). Adiabaticity can easily be achieved, and the transmission is steady. Because this fiber initially also supports the $LP_{02}$ and $LP_{21}$ modes and because our launch can weakly excite these modes, we observe a slight drop in transmitted power near $R = 20~\mu$m, when these higher-order modes become cladding modes. Regime $B$ occurs after the light has escaped from the core to the cladding. Because the cladding is typically much larger than the core ($R/a~>~10$), and because $n_{clad} - n_{air} \gg n_{core}-n_{clad}$, $V$ increases by over two orders of magnitude ($V_{clad} \approx 200$). If the core-cladding transition is not adiabatic, modes of similar symmetry are excited. In particular, we observe transfers of energy to the $LP_{12}$ family which contains the modes $TE_{02}$, $TM_{02}$, and $HE_{22}$. The interaction between those modes results in mode beating inside the fiber, and an oscillation in the amount of output light. The oscillations continue to $R\approx0.7~\mu$m (regime $C$), where for a typical silica fiber, $V_{clad} \approx 6$. From this cutoff location, it is clear that much of the beating behavior is due to the $LP_{12}$ modes which have been excited through non-adiabatic transitions. The pull extends through the $TM_{01}$, $TE_{01}$, $HE_{21}$ cutoffs (regime $D$). After reaching the cutoff radius near $R=290$ nm, very little light reaches the photodetector - typically less than 1-2\% - indicating low population of the fundamental mode. To determine the loss in the $HE_{11}$ mode, we measured the transmitted power before and after the pull by translating the phase step out of the way. The losses in this mode were negligible, meaning that we reach a mode purity of 98-99\%. Mode purity is of fundamental importance for nanofiber applications, and in particular for atomic physics where polarization control and stability are required.

We have compared the transmissions obtained using fibers of diameter 80~$\mu$m and 50~$\mu$m, and observed the beneficial effects of smaller clad fibers and higher numerical aperture for $\Omega=2$ mrad. This taper angle is chosen because it is non-adiabatic and highlights the effect of fiber diameter on adiabaticity. The improvement in adiabaticity is clear in Fig.~\ref{fibertype}, with a transmission of 52\% obtained for the 50-$\mu$m fiber, compared to the 10\% transmission of the 80-$\mu$m fiber. The table in Fig.~\ref{fibertype} compares the properties of the 80-$\mu$m and 50-$\mu$m fibers. Note that the SM1500 fiber initially supports the $LP_{01}$, $LP_{11}$ as well as the next families of excited modes $LP_{02}$ and $LP_{21}$ ($V_{core,R_0}=4.3$). These higher-order modes are substantially filtered, though not completely, by winding the fiber around a 4 mm-diameter rod. 

Using the numerical aperture (NA) and the cutoff wavelength provided by Thorlabs and Fibercore, we are able to derive using Eq.~(\ref{V}) the initial $V$ value $V_{core,R_0}$, the radius at which the $LP_{11}$ family escapes from the core to the cladding $R_{c}^{LP_{11}}$ ($V_{core,R_c}=2.405$: the mode enters the cladding-to-air guidance regime) and the corresponding fiber $V$-number at that radius $V_{clad,R_c}$ for a wavelength of 780 nm. Empirically, we obtain an estimate of $R_{c}^{LP_{11}}$ on the transmission plot when the oscillations start. The numbers agree quantitatively. We see that reducing the initial cladding radius and increasing the fiber numerical aperture allows the $LP_{11}$ family to leak from the core to the cladding at a smaller fiber radius. This results in a significantly reduced $V_{clad}$ when the modes escape from the core to the cladding: By reducing the cladding diameter from 80~$\mu$m to 50~$\mu$m and increasing the numerical aperture from 0.18 to 0.30, the number of available modes decreases by more than an order of magnitude for these two commercially available fibers. Improvements could be achieved by pre-etching the fiber to a smaller diameter so that the initial core radius to cladding radius ratio $R/a$ is further reduced. In this case, the numerical aperture remains unchanged, and the number of available modes directly scales with the square of the ratio of initial cladding radii.

\subsubsection{Spectrograms}

In non-adiabatic propagation, the $LP_{11}$ modes couple to higher-order modes of the same symmetry, belonging to families $LP_{1m}$ ($m\geq2$). Because they propagate with different propagation constants during the pull, they accumulate a phase difference leading to interference in the amount of light that coupled back into the core. Since the photodetector only measures core light, this interference leads to oscillations in the transmission (Fig.~\ref{fibertype}). A useful way to examine these data is through spectrograms, which plot local, windowed Fourier transforms of the transmission signals as a function of waist radius. Plotting the spectrogram of the transmission signal \cite{Ravets2013,Orucevic2007}, we directly observe the contribution of various pairs of modes in the beating. Figure~\ref{fibertype} shows the spectrograms for the 80-$\mu$m and 50-$\mu$m diameter fiber pulls. Each line in the spectrogram is specific to the beating between a $LP_{01}$ or $LP_{11}$ mode and another mode of similar symmetry excited during the pull. The curve ends when one of the modes reaches its cutoff: the energy is then lost via coupling to radiative modes.

The number of lines observed in a spectrogram is directly related to the excitation of higher-order modes through non-adiabatic processes from a single launched mode: the more lines present in the spectrogram, the less adiabatic the pull is. Moreover, the colormaps in Fig.~\ref{fibertype} are normalized in such a way that the intensity of each red line gives the strength of the energy transferred. It is clear from Fig.~\ref{fibertype} that more intense lines are present in the 80-$\mu$m fiber than in the 50-$\mu$m, further supporting our observation of more stringent adiabaticity requirements for fibers with a large cladding radius and small numerical aperture.

\subsubsection{Imaging the fiber output}

We monitor the transmitted beam with a CCD, and obtain Movie 2 showing both the core and cladding light as a function of time throughout the tapering process. Using a microscope objective, we first image the end of the fiber to observe the core-guided light (Movie 1 in Fig.~\ref{fibertype}). The movie shows oscillations that result from mode competition, which modulates the amount of light that exits the fiber in the core.

\begin{figure}[H] 
\centering
\includegraphics[width=10cm]{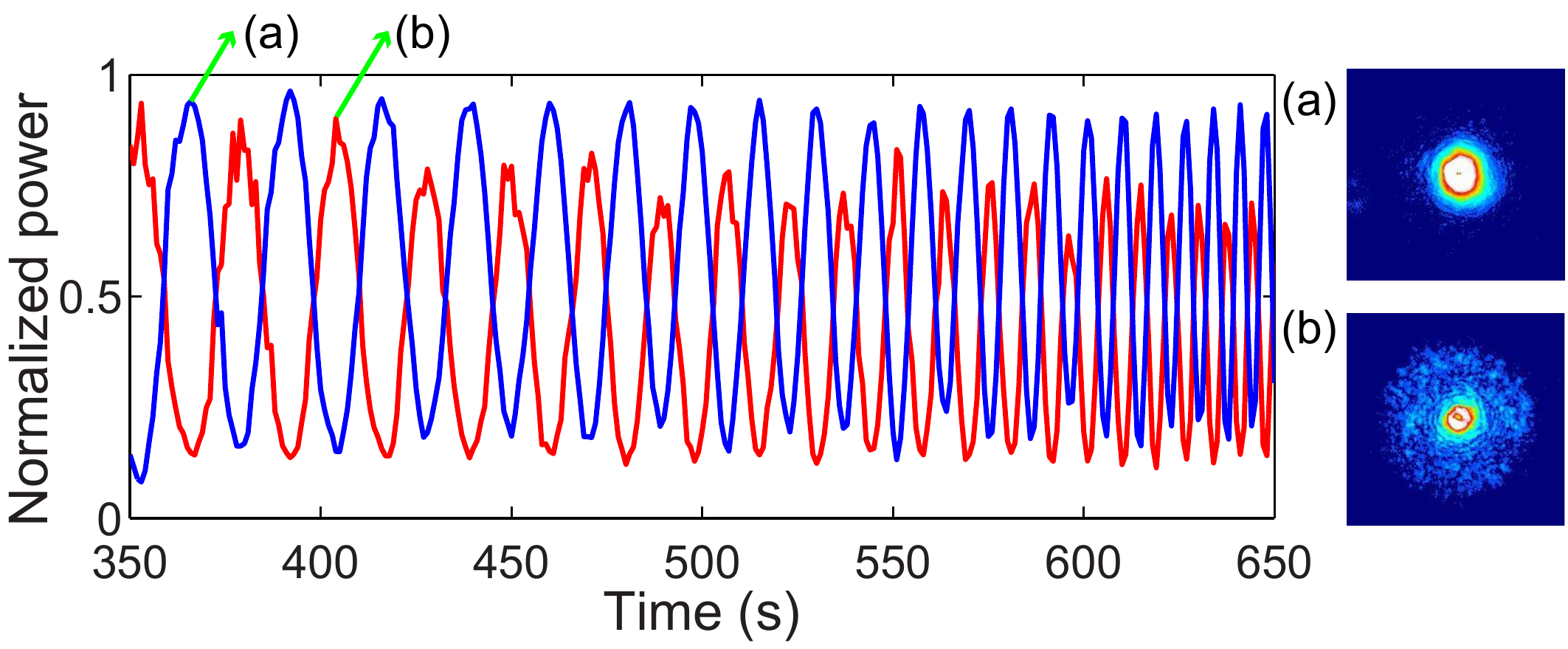}
\caption{Amount of light (normalized) exiting the fiber from the core (blue curve) and from the cladding (red curve). The signals are out of phase, confirming the transfer of energy between modes during the tapering. We observed the two simultaneously by using the two reflections from a thick beamsplitter. (a)-(b) (1.1 MB) Movie 2 shows the evolution of the beam transmitted through a nanofiber during a portion of a pull, where the power is high enough to observe the cladding light.}
\label{corecladding}
\end{figure}

Movie 2 in Fig.~\ref{corecladding}(a) also depicts the transfer of energy from the core to the cladding. Because the cladding intensity is low compared with the core, we record two spatially separated images of the fiber simultaneously by using strong and weak reflections from a thick beamsplitter with AR-coated front face. Figure~\ref{corecladding} shows the normalized fractions of energy exiting the fiber from the cladding and from the core. The two signals are out of phase. We note that the sum of the energy contained in the core and the energy contained in the cladding does not add up to the total energy input in the fiber. Outside the taper region, cladding light becomes highly scrambled and lost through the fiber buffer, resulting in both the speckle observed in Movie 2 and the reduced total transmitted power. Residual cladding light is spatially filtered from hitting the photodetector, so that the observed oscillations Fig.~\ref{fibertype} are due only to core-guided light.

\subsection{Varying the angle}

\begin{figure}[H]
\centering
\includegraphics[width=10cm]{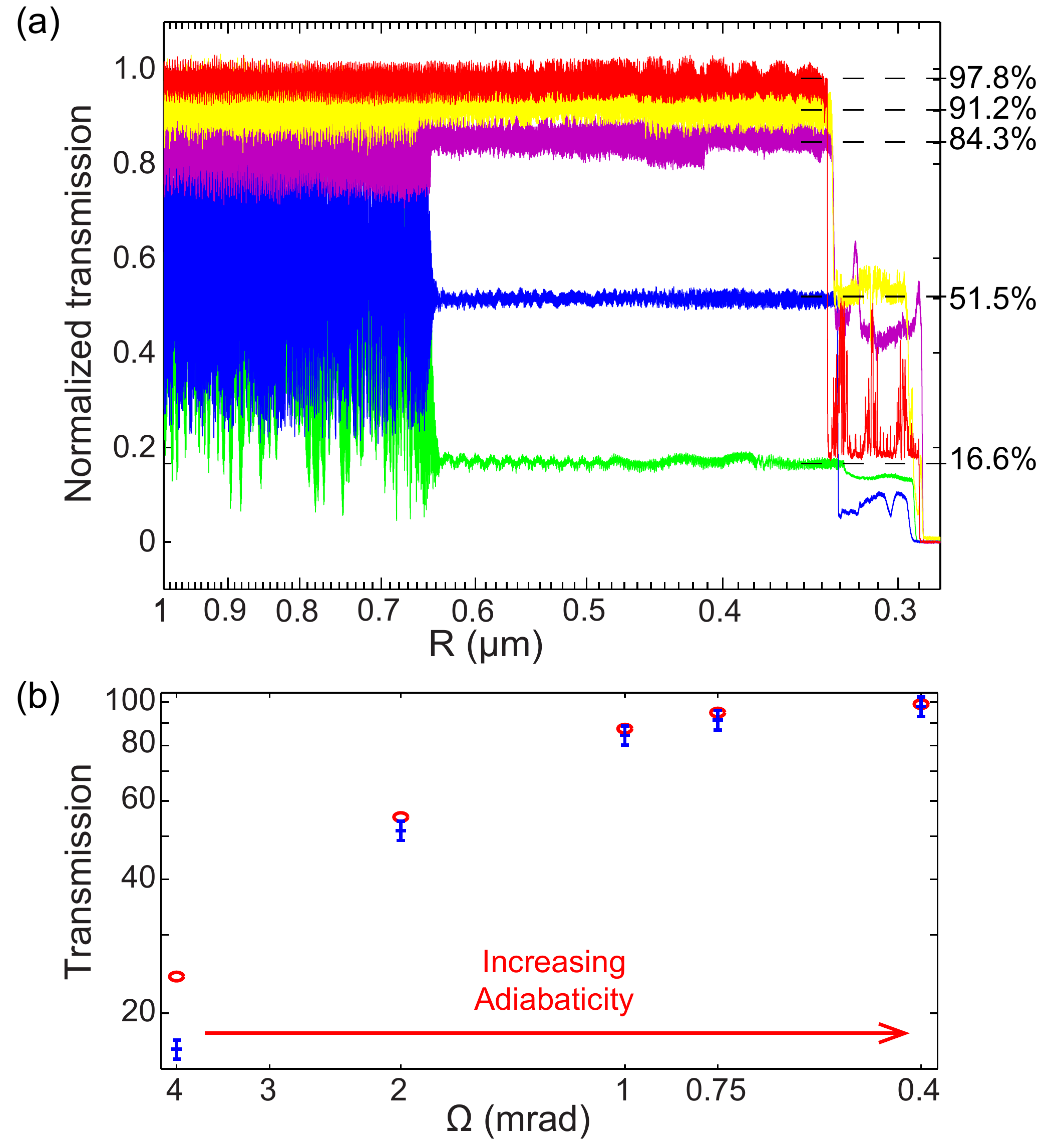}
\caption{(a) Transmission through a SM1500 nanofiber for $\Omega$ = 4~mrad (green), 2~mrad (blue), 1~mrad (purple), 0.75~mrad (yellow), 0.4~mrad (red). The transmission plots have the fundamental mode subtracted out (typically about 1\%) to accurately describe the total transmission of the $LP_{11}$ modes. The horizontal axis for the 4 mrad pull was renormalized to take into account extra tension in the fiber due to the rapidity of the pull. (b) Simulated (red circles) and experimental (blue lines) final transmissions through the fiber as a function of angle. Decreasing $\Omega$ enables us to improve the transmission of the $LP_{11}$ family up to 97.8\%.}
\label{multiplepulls} 
\end{figure}

The measurements in this section use the reduced-cladding Fibercore SM1500. For this fiber, the $LP_{11}$ modes transition to cladding modes near $R$ = 13 $\mu$m. By using a reduced-diameter fiber, we observed a drastic improvement of the transmission of the $LP_{11}$ modes. To further improve the adiabaticity, it is necessary to look into more details of the tapering process itself. We have lowered $\Omega$ to improve the transmission over what is observed in the previous section. Figure~\ref{multiplepulls} shows the results of draws using $\Omega =$ 4, 2, 1, 0.75 and 0.4 mrad. Although each plot shows the same qualitative behavior as described in Fig.~\ref{fibertype} for the 2-mrad pull, the strength of the features depends on $\Omega$.

The free-space mode at the fiber input has a spatial polarization that is an equal superposition of $HE_{21}$ and $TM_{01}$ (or $TE_{01}$). However, mode conversion occurs where the fiber is wound around the mode-filtering mandrel so that the distribution entering the nanofiber is unknown. Within the nanofiber waist, which is held fixed and straight, mode conversion is unlikely to occur so that the desired $LP_{11}$ mode can be achieved after the pull~\cite{Fatemi2011}. By $R = 0.45~\mu$m, only the $TM_{01}$, $TE_{01}$, and $HE_{21}$ modes are confined, with a small contribution in the fundamental $HE_{11}$. The $HE_{12}$ mode achieves cutoff at $V_{clad}=2.8$ ($R \approx 330$ nm), earlier than the $TM_{01}$ and $TE_{01}$ modes, which reach cutoff at $V=2.4$ ($R \approx 290$ nm). At $R \approx 330$ nm, the power is reduced by the $HE_{21}$ content, which is determined by the initial superposition of states entering the nanofiber.

The transmitted power in the $LP_{11}$ mode family is 16.6\% for $\Omega = 4$ mrad. The transmitted power drops sharply near $R$ = 13~$\mu$m, and undergoes strong oscillations between 10-80\%. For $\Omega = 2$ mrad, the oscillations below $R=13~\mu$m are reduced, with the transmission fluctuating between 30-90\%, and the transmitted power improves to 51.5\%. Further decreasing the angle to $\Omega =$ 1, 0.75 and 0.4 mrad, improves the transmission to 84.3\%, 91.2\% and 97.8\% respectively, where uncertainty is dominated by systematic effects that should be less than 1\%. Figure~\ref{multiplepulls} also shows excellent agreement between the experimental results and those obtained using commercial waveguide propagation software~\cite{FIMMPROP}. For those calculations, the propagation was modeled using the targeted fiber geometry. In our transmission measurements, the scaling between time and radius is made using a separate algorithm that models the dynamics of the pull based on conservation of volume \cite{Hoffman2013}. For $\Omega =$ 0.4 to 2 mrad, the mode cutoff positions we obtain using the scaling from the algorithm correspond directly to what is expected theoretically. We also observe excellent agreement between the experimental transmission measurements and the calculations, confirming the accuracy of our pulling procedure. We were not limited by systematic effects that might include accumulation of contaminants for longer pulls, and asymmetric profiles for faster ones. For the fastest pull (4 mrad), we observed a discrepancy due to systematic effects, and we must apply a different scaling to match the theoretical cutoffs.

For $\Omega=0.4$ mrad, the most shallow angle studied, the amplitude of the oscillations, which is directly related to the energy transfer to undesired modes, is reduced to a few percent. The observed transmissions are due to losses into and out of the waist. Because the fiber is symmetric, the normalized transmitted power is the square of that in the waist. For $\Omega = 0.4$ mrad, this leads to 98.9\% power in the waist. We believe that such a fiber is usable for various applications involving higher-order modes. We note that reaching adiabaticity for higher-order modes requires fibers that are substantially longer than for the fundamental mode. For the $HE_{11}$ mode, transmissions greater than 99\% can easily be achieved for $\Omega$ up to 5 mrad \cite{Ravets2013}. We also observed that when we remove the $\pi$-phase plate from the launch, the transmission in the fundamental mode is essentially equal to the transmission before pulling, as adiabaticity is strongly satisfied for this mode. During the tapering process, the higher-order modes escape from the core to the cladding earlier than the fundamental $HE_{11}$ mode. When the $HE_{11}$ mode finally transitions to a cladding mode, $R$ has decreased, so that $V$ and the number of available modes to couple to is smaller. The reduction in $R$ also leads to an increase in the difference between adjacent propagation constants, allowing less mode interaction and a steeper $\Omega$. Mode conversion also occurs throughout the fiber and not just at the core-cladding transition point, but this region has the most stringent adiabatic criterion.

\section{CONCLUSION}

We have demonstrated propagation of higher-order modes in nanofibers using the $TE_{01}$, $TM_{01}$, and $HE_{21}$ modes. By tapering the fiber with angles near 0.4 mrad and using a commercial, off-the-shelf fiber with 50~$\mu$m diameter, we have achieved transmission efficiency of 97.8\% with excellent mode purity, a factor of four higher than previous work, and more than one order of magnitude improvement on mode purity. Critical to this work was a spectrogram analysis of the modes present during the pulling. Our experimental results agree with simulations of the propagation through the taper. High transmissions of $LP_{11}$ modes with high purity is a promising tool for atomic physics, expanding the possible intensity and polarization configurations of evanescent fields surrounding the nanofiber.

\section*{ACKNOWLEDGEMENTS}

This work was funded by ONR, the ARO Atomtronics MURI, DARPA, and the NSF through the PFC at JQI. S.~R. acknowledges support from the Fulbright Foundation.


\end{document}